\documentclass{aa}  

\usepackage{hyperref}

\usepackage{natbib}
\bibpunct{(}{)}{;}{a}{}{,}

\usepackage[varg]{txfonts}
\usepackage{epsf,palatino,url,graphicx,wrapfig,array,setspace,amsmath,amssymb,fancyhdr,multirow,lscape,appendix,rotating}
\usepackage{graphics,epsfig}

\usepackage{longtable}
\usepackage{amssymb}
\usepackage{xcolor}
\usepackage{color}
\usepackage{lipsum}
\usepackage{textcomp}
\usepackage{threeparttable}
\usepackage{enumerate}

\usepackage{placeins}
\usepackage{float}
\usepackage{tabularx}

\usepackage[normalem]{ulem}
\usepackage{soul,color}
\usepackage{blindtext}

\newcommand{\ergs}[1]{$\times 10^{#1}$ erg s$^{-1}$}

\newcommand{\hcm}[1]{$\times 10^{#1}$ cm$^{-2}$}
\newcommand{\ohcm}[1]{$10^{#1}$ cm$^{-2}$}

\newcommand{\ltsima}{$\buildrel < \over \sim$}
\newcommand{\lsim}{\lower.5ex\hbox{\ltsima}}
\newcommand{\gtsima}{$\buildrel > \over \sim$}
\newcommand{\gsim}{\lower.5ex\hbox{\gtsima}}

\newcommand{\swift}{{\it Swift}\xspace}

\newcommand{\xmm}{{\it XMM-Newton}\xspace}
\newcommand{\nus}{{\it NuSTAR}\xspace}

\newcommand{\cxo}{\hbox{Chandra}\xspace}

\newcommand{\ulx}{NGC\,300\,ULX1\xspace}

\newcommand{\eqb}{\begin{eqnarray}}
\newcommand{\eqe}{\end{eqnarray}}

\begin{document} 

\defcitealias{2018MNRAS.476L..45C}{C18}

\title{NGC 300 ULX1: A test case for accretion torque theory}

\author{G. Vasilopoulos\inst{1,2}\thanks{\email{georgios.vasilopoulos@yale.edu}}
\and F. Haberl\inst{1} 
\and S. Carpano\inst{1} 
 \and C. Maitra\inst{1}} 

\titlerunning{Temporal properties of NGC300 ULX1}
\authorrunning{Vasilopoulos et al.}

\institute{Max-Planck-Institut für Extraterrestrische Physik, Giessenbachstra{\ss}e, 85748 Garching, Germany \and Department of Astronomy, Yale University, PO Box 208101, New Haven CT 06520-8101, USA}

\date{Received ... / Accepted ...}

\abstract
     {NGC 300 ULX1 is a newly identified ultra-luminous X-ray pulsar. The system is associated with the supernova impostor SN 2010da that was later classified as a possible supergiant Be X-ray binary. In this work we report on the spin period evolution of the neutron star based on all the currently available X-ray observations of the system. 
     We argue that the X-ray luminosity of the system has remained almost constant since 2010, at a level above ten times the Eddington limit. 
     Moreover, we find evidence that the spin period of the neutron star evolved from $\sim$126\,s down to $\sim$18\,s within a period of about 4 years.  
     We explain this unprecedented spin evolution in terms of the standard accretion torque theory.
     An intriguing consequence for NGC 300 ULX1 is that a neutron star spin reversal should have occurred a few years after the SN 2010da event.  
}


\maketitle


\section{Introduction}
\label{sec:intro}

X-ray binaries are among the most luminous stellar-mass objects that are powered by accretion.
A major implication of spherical accretion is that the released radiation can become so luminous that in principle it could halt accretion (at the Eddington limit L$_{\rm Edd}$).
Nevertheless, numerous sources have been observed at X-ray luminosity (L$_X$) levels well above L$_{\rm Edd}$ for a neutron star (NS).
These are the so-called  ultra-luminous X-ray sources (ULXs) \citep{2017ARA&A..55..303K}.
ULXs are considered promising candidates for hosting heavy stellar-mass black holes.
However, the X-ray spectral properties of many ULXs are inconsistent with sub-Eddington accretion models, implying super-Eddington accretion onto stellar-mass objects \citep{2016AN....337..534R}.
Remarkably, over the last years pulsations have been discovered from a few such systems \citep{2014Natur.514..202B,2016ApJ...831L..14F,2017MNRAS.466L..48I,2017Sci...355..817I}.
This offers undisputed evidence that at least a few ULXs host highly magnetized NSs.
Recently, \citet{2017A&A...608A..47K} revisited a large sample of known ULXs and by taking into account theoretical predictions \citep[e.g.][]{2017MNRAS.468L..59K,2017MNRAS.467.1202M} argued that a significant fraction of non-pulsating ULXs may as well be powered by a highly magnetized NS.

\ulx is a newly identified ULX pulsar (ULXP) \citep[][hereafter C18]{2018MNRAS.476L..45C}, located in NGC 300 at a distance of 1.88 Mpc \citep{2005ApJ...628..695G}.
The system became active in X-rays and optical in May 2010, when its luminosity rapidly increased causing it to exhibit what was classified as a supernova impostor event \citep[SN 2010da,][]{2011ApJ...739L..51B,2016ApJ...830..142L}.
In \citetalias{2018MNRAS.476L..45C} we showed that, in the early 2010 observations the spectrum of the system was mostly affected by partial absorption (i.e. equivalent hydrogen column density $N_H\sim$5\hcm{23}), while in the 2016 spectrum the $N_H$ of the partial absorption component was significantly lower by a factor of $\sim$100.
From the analysis of \xmm and \nus data we derived an unabsorbed X-ray luminosity ($L_X$) of 4.7\ergs{39} (0.3-30.0 keV), a value consistent with both 2010 and 2016 observations.
The temporal analysis of the 2016 X-ray data revealed a spin period ($P$) of 31.7~s and a spin period derivative ($\dot{P}$) of -5.56$\times$10$^{-7}$ s/s.  
In this letter, we  discuss the NS spin-period evolution in terms of standard accretion theories and we will comment on its derived properties that make \ulx such an exciting case study among ULXPs.


\section{Observational data and analysis}
\label{sec:observations}

Since its X-ray brightening in 2010, \ulx has been monitored by most of the modern X-ray observatories; \swift, \xmm, \nus and \cxo. 
We requested additional \swift monitoring and two \cxo target of opportunity (ToO) observations in order to measure the most recent spin-up rate and $L_X$ of the system.  
Additionally, a \nus ToO observation was performed on January 2018 \citep{2018ATel11282....1B}.
For all available X-ray observations standard products were extracted using the latest available software packages and instrument calibration files.
For \xmm  data reduction we used SAS v16.1.0. \nus and \swift/XRT data were reduced using {\tt nupipeline} (v0.4.6) and {\tt xrtpipeline} (v0.13.4) respectively, that can be found in  HEASoft 6.22 software \citepalias[see details in][]{2018MNRAS.476L..45C}. 
\cxo data reduction was performed with {\tt CIAO} v4.9 software \citep{2006SPIE.6270E..1VF}.

The X-ray spectra were analyzed using the {\tt xspec} (v12.9.0) spectral fitting package \citep{1996ASPC..101...17A}.
For the fits we used a phenomenological model that best described the high quality X-ray spectra obtained during three \xmm observations performed in 2010 and 2016 \citepalias{2018MNRAS.476L..45C}. 
The continuum model consists of a power law with a high-energy exponential cutoff ($E_{\rm cut}\sim$6 keV) and a disk black-body component that accounts for emission from the accretion disc.
Absorption was modeled by a partial coverage component and an absorption component to account for the local and interstellar absorption, respectively.
For each of the analyzed spectra we used a Bayesian approach to derive the probability density distributions of the marginalized parameters of the model and the intrinsic $L_X$ of \ulx.  
Additional details about the spectral fitting are provided in Appendix~\ref{sec:spec}.

\begin{figure}
  \resizebox{\hsize}{!}{
  \includegraphics[angle=0,clip=]{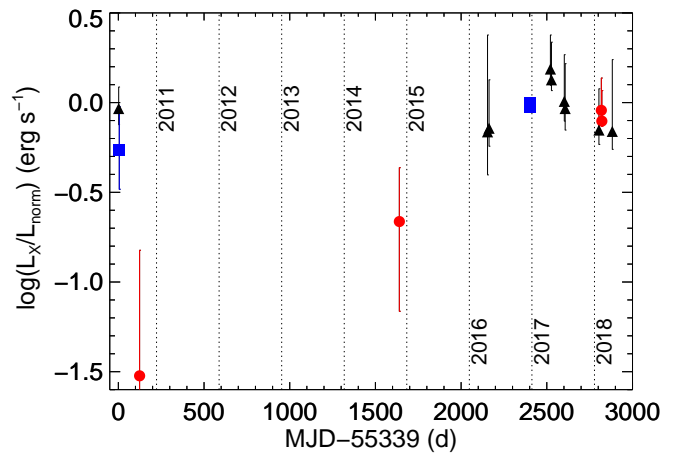}
 }
  \vspace{-0.5cm}
  \caption{Normalized $L_X$ of \ulx, as derived by the fits to the X-ray spectra from \swift (black triangles), \xmm (blue squares) and \cxo (red  circles). Values are scaled to 4.7\ergs{39}, the $L_X$ derived by  \citetalias{2018MNRAS.476L..45C} (see text).} 
  \label{fig:lc}
\end{figure}

The X-ray light curve of \ulx is shown in Fig. \ref{fig:lc}. For simplicity we 
scaled the $L_X$ obtained from our spectral analysis (0.3-10.0 keV) 
to the $L_X$ derived from the simultaneous NuSTAR and XMM-Newton observations 
in 2016 \citepalias[$L_X=$4.7\ergs{39};][]{2018MNRAS.476L..45C}. 
Between 2011 and 2016, only few observations were performed with collected events consistent 
with heavily absorbed spectra \citepalias[see also][]{2018MNRAS.476L..45C}. 
Due to the low statistics in these observations there is a large degeneracy between the model parameters and especially the disk black body component cannot be constrained. 

To further investigate X-ray variability we have simultaneously 
fitted all the spectra with low statistics with the same continuum model and 
letting only the parameters of the absorption components
vary for each spectrum. Fixing $L_X$ to 4.7\ergs{39}
and the continuum parameters to the best fit parameters
inferred from the 2016 \xmm observation \citepalias[see Table 1 in][]{2018MNRAS.476L..45C}, yielded an acceptable fit for all
spectra. 
Thus, between 2010 and 2018 the observed variability is consistent with $L_X$ changes no more than a factor of three.

We note that our aim was not to investigate whether our phenomenological 
model describes the data best, but to put constraints on the $L_X$ evolution 
over the $\sim$8 y period following the SN 2010da event. 
For ULXPs a plethora of models has been used to describe
their spectra; some motivated by X-ray pulsars in
high mass X-ray binaries \citep[e.g. ][]{2018ApJ...857L...3W}, or in the
context of optically thick accretion envelopes that has been
proposed to explain ULXP spectra \citep{2017A&A...608A..47K,2017MNRAS.467.1202M}. 
A detailed study of all these models
is beyond the scope of this letter and will be further
addressed in an upcoming work \citep{2018arXiv181106251K}.

\begin{figure}
  \resizebox{\hsize}{!}{
  \includegraphics[angle=0,clip=]{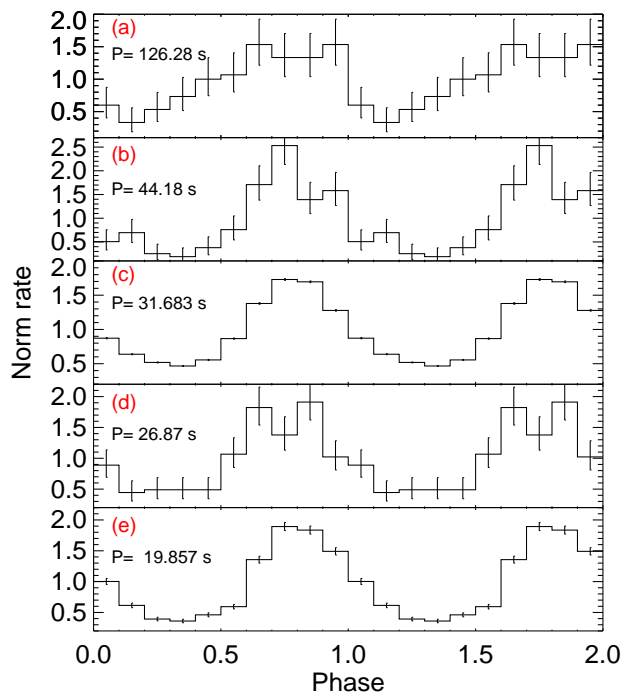}
 }
  \vspace{-0.5cm}
  \caption{Normalized pulse profiles of \ulx. Subplots correspond to different epochs (see Table \ref{tab:log}) and are sorted in decreasing spin period (top to bottom). 
  Pulse profiles derived by \cxo (panels a and e) contain events with energies in the 0.3-8.0 keV band, while those derived from \xmm data (panel c) and \swift (panels b and d) denote the 0.3-10.0 keV range. 
  } 
  \label{fig:pp}
\end{figure}

Given the extreme spin-up rate of the NS in \ulx reported by \citetalias{2018MNRAS.476L..45C}, $P$ can significantly change within the duration of a typical X-ray exposure ($>$10 ks).  
To take into account this effect we performed an accelerated epoch folding (AEF) test \citep{1983ApJ...272..256L} to derive the ephemeris of \ulx.
Details about this methodology and its caveats can be found in Appendix~\ref{sec:temp}.
The complete list of observations and the derived properties of the system are listed in Table \ref{tab:log}. 
A periodic signal was detected for all observations performed after November 2014, when a period of $\sim$126.3~s was detected on MJD 56978. 
The latest analyzed observation yielded a period of $\sim$19~s on MJD 58221 (see Fig. \ref{fig:pp}).

The \swift/XRT monitoring of \ulx is ongoing.
These additional observations cover the period after May 2018 and will be presented in a future study \citep[see also NICER monitoring][]{2018arXiv181109218R}.
However, we have performed a preliminary spectral and temporal analysis 
and have measured a periodic signal in all of them.
The \swift/XRT observations show that $L_X$ is still consistent with being almost constant within a factor $<2$ till about August 2018.
The latest detected spin period of \ulx is $\sim$17.977~s measured on MJD 58307.3.

\section{Results \& Discussion}
\label{results}

We first describe briefly the principles of the interaction between the NS and accretion disk around it \citep[see also review of][]{2014EPJWC..6401001L},
and then we apply them to the observational data of \ulx. Throughout the calculations we will assume typical values of $M_{NS}=1.4M\sun$ and $R_{NS}=10$~km for the NS mass and radius, respectively.

If the inner radius of the Keplerian disk ($R_{d}$) is inside the co-rotation radius ($R_{co}$), then accretion along the magnetic lines can occur.
If the $R_{d}>R_{co}$, accretion is halted by a centrifugal barrier \citep[i.e. propeller regime;][]{2018A&A...610A..46C}.
During accretion, we can observationally measure the $L_X$ emitted by the accreted material and the change of the NS $P$.
In general, $L_X$ is equal to the rate at which gravitational energy of the in-falling matter is released. 
Nevertheless, the fraction of the gas gravitational potential energy converted to radiation could be different \citep[e.g. radiative inefficient flow; ][]{2015MNRAS.449.2803D} and model dependent for the environment of the accretion column \citep{2016A&A...591A..29F,2017ApJ...835..129W}. 
Thus we can define as $n_\mathrm{eff}$ the efficiency that gravitational energy is converted to radiation ($L_{x}={n_\mathrm{eff}{GM_{NS}}\dot{M}/R_{NS}}$).
To calculate the induced torque onto the NS, we further assume that the accretion disk is truncated at the magnetospheric radius $R_{M}$ due to the strong magnetic field $B$ of the NS:
\begin{equation}
R_{M} = \xi \left(\frac{R_{NS}^{12}B^4}{2GM_{NS}\dot{M}^2}\right)^{1/7},
\label{eq0}
\end{equation}
where the $\xi$ parameter takes a value of $\sim$0.5 \citep{2018A&A...610A..46C}.

Several models have been proposed for calculating the applied torque onto an accreting NS \citep[e.g.][]{1987A&A...183..257W,1979ApJ...234..296G,2016ApJ...822...33P}.
In the early paradigm presented by \citet{1979ApJ...234..296G}, the magnetic field lines permeate through the disk and couple the latter with the NS.
Due to the mismatch of the angular velocity of the disk with the NS angular velocity, lines that intersect the disk inside (outside) $R_{co}$ apply a positive (negative) torque to the NS. 
The torque due to mass accretion is $N_{acc}\approx\dot{M}\sqrt{GMR_{M}}$. 
However, the total torque can be expressed in the form of $N_{tot}=n(\omega_\mathrm{fast})N_{acc}$ where $n(\omega_\mathrm{fast})$ is a dimensionless function of $\omega_\mathrm{fast}=(R_{M}/R_{co})^{3/2}$, which is known as the fastness parameter. 
Assuming the magnetic field lines thread through the disk an assessment of the magnetic stresses \citep{1995ApJ...449L.153W} yields:    
\begin{equation}
n(\omega_\mathrm{fast})=\left( 7/6-(4/3)\omega_\mathrm{fast}+(1/9)\omega_\mathrm{fast}^2 \right) / \left( 1-\omega_\mathrm{fast} \right),
\label{eq1}
\end{equation}
which simplifies to $n(\omega_\mathrm{fast})=7/6$ for very slow rotators ($\omega_\mathrm{fast}<<1$).
Alternatively, a more physical picture is that, the magnetic field lines cannot remain connected to the NS and the spin-down term is due to the enhanced opening of magnetic field lines \citep[see][]{2016ApJ...822...33P}.
This result is also supported by simulations \citep{2017MNRAS.469.3656P,2017ApJ...851L..34P}.
For the range of the derived parameter values of \ulx (i.e. $P$, $L_X$, $B$), the spin-down due to enhanced opening of magnetic field lines is more than two orders of magnitude smaller than the spin-up due to accretion.
Thus, one can safely assume that $N_{tot}=N_{acc}$. 
We note that all the treatments mentioned above vary only by a normalization factor when applied to ``slow rotator'' systems away from the propeller transition (i.e. $\omega_\mathrm{fast}<<1$).
In simple terms, for $n(\omega_\mathrm{fast})=1$ the formulation of \citet{1979ApJ...234..296G} produces the same results as those of \citet{2016ApJ...822...33P}.

\begin{figure}
  \resizebox{\hsize}{!}{
\includegraphics[angle=0,clip=]{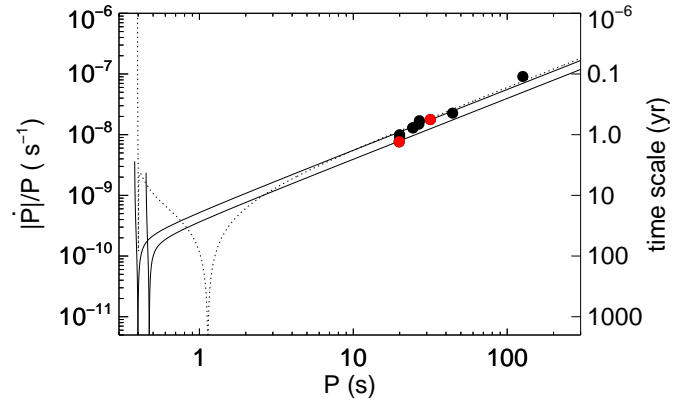}
 }
  \vspace{-0.5cm}
   \caption{
   $|\dot{P}|/P$ vs. $P$ for \ulx.
   Solid black lines denote the solution using the W95 approximation while assuming a constant $L_X$. 
   The two solutions based on the derived values from the two data-sets (red points) with best statistics (2016 \& 2018) presented in Table \ref{tab:mab}. 
   The \citet{1979ApJ...234..296G} model prediction is plotted (dotted line) for comparison using the observed 2016 $L_X$.
   All model lines assume $n_{\rm eff}=100\%$.}
  \label{fig:PPdot_B}
\end{figure}

Having a model of the accretion torque we can then compute the spin derivative $\dot{P}$ of the NS.
For a NS with moment of inertia $I_{NS}=2M_{NS}R_{NS}^2/5$ one derives:
\begin{equation}
-\frac{\dot{P}}{P}=\frac{P}{I_{NS}{\rm 2\pi}} n(\omega_\mathrm{fast}) \dot{M} \sqrt{G M_{NS} R_{M}},
\label{eq2}
\end{equation}
from where we can solve for the polar magnetic field strength:
\begin{equation}
B \simeq 30 n(\omega_\mathrm{fast})^{-7/2} \xi^{-7/4} G^{3/2} (\dot{P}/P^2)^{7/2} M_{NS}^5 R_{NS} n_\mathrm{eff}^3 L_X^{-3}
\label{eq3}
\end{equation}
We applied eq. \ref{eq3} to observations with good statistics where $P$ and $\dot{P}$ were computed with high accuracy (i.e. only two epochs). 
We estimated $B$ to be approximately $10^{12-13}$~G (see Table \ref{tab:mab}).
Interestingly, for both epochs we derive the same value for $B$, which in turn implies that $\xi$ and $n_\mathrm{eff}$ have not changed.
The high magnetic field strength is consistent with the findings of \citet{2018arXiv180605784T}.
By using the values of Table \ref{tab:mab}, we can derive the $P$-$\dot{P}$ evolution of the system assuming constant accretion, as is shown in Fig.\,\ref{fig:PPdot_B}. 
\ulx is clearly still away from the equilibrium (i.e. $\dot{P}=0$); which when reached will be less than 1\,s, assuming constant $\dot{M}$.

\begin{table}
\begin{center}
\resizebox{\columnwidth}{!}{
\begin{threeparttable}
\begin{tabular}{lcccccc}
\hline\noalign{\smallskip}  
 Date$^{(a)}$ & $P$ & $\log(|\dot{P}|)$ & $n_\mathrm{eff}$ & $L_X/L_\mathrm{norm}$$^{(b)}$    & B (W95) & B (P16) \\
   (y/MJD)    & (s) & (s/s)  &  $\%$ & erg/s  & \multicolumn{2}{c}{$10^{12}$ G}       \\
\hline\noalign{\smallskip}    
2016/57738 & 31.718 & -6.257 & 100\% & 1  & 5.8 & 9.9 \\
   & 31.718 & -6.257 & 50\% & 1  & 0.7 & 1.2 \\
2018/58149 & 19.976 & -6.74  & 100\% & 0.8  & 5.9& 10.0\\
   & 19.976 & -6.74  & 50\% & 0.8  & 0.7 & 1.3\\     
\noalign{\smallskip}\hline\noalign{\smallskip}

\hline 
\end{tabular}
{
 \tnote{(a)} For this table we used only the data from the 2016 and 2018 periods when the system was observed by \nus. 
 \tnote{(b)} $L_X$ is normalized to 4.7\ergs{39}, as derived by \citetalias{2018MNRAS.476L..45C}.  }
\end{threeparttable}
 }
\end{center}
\caption{Estimation of $B$ under different assumptions for the \citet{1995ApJ...449L.153W} (W95) and \citet{2016ApJ...822...33P} (P16) torque models.}
\label{tab:mab}
\end{table}

\begin{figure}
  \resizebox{\hsize}{!}{
     \includegraphics[angle=0,clip=]{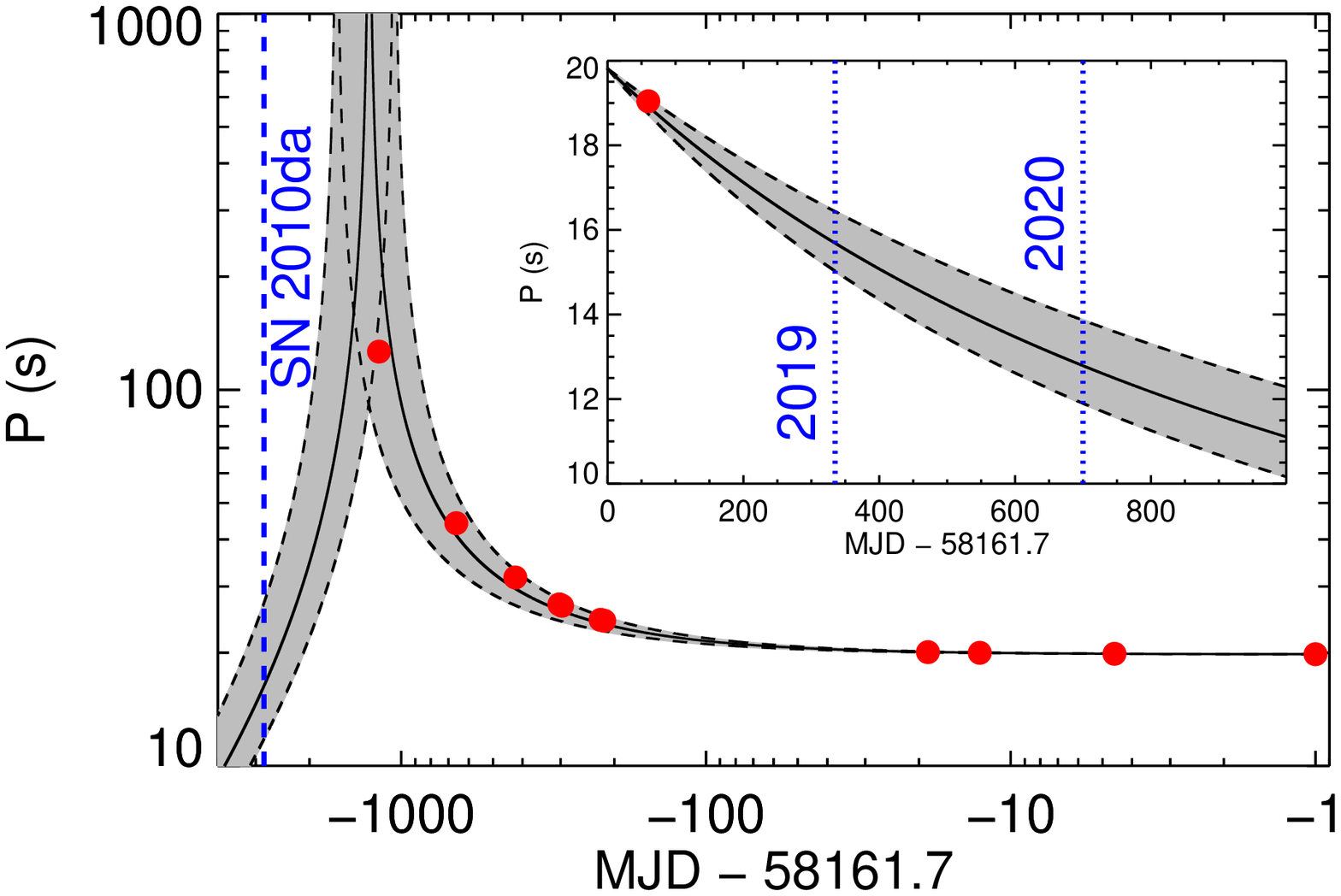}
     }
  \vspace{-0.5cm}
  \caption{Evolution of the NS spin period of \ulx since the SN 2010da event (blue vertical dashed line). The black solid line denotes the theoretical prediction assuming a constant accretion rate and stellar torque from disk-magnetosphere interactions \citep{1995ApJ...449L.153W} with efficiency = 1.0, $B$ = 5$\times$10$^{12}$~G and $L_x$ = 4\ergs{39}. The gray shaded area marks the evolution path for a range of $L_X$, i.e. 3--5\ergs{39}. The inset plot shows the predicted evolution of the spin period for the next years, with the lone data point being the $\sim19$~s \swift measurement (MJD 58221).}
  \label{fig:evolP}
\end{figure}

Having an estimate of $B$ we can numerically solve eq. \ref{eq2} backwards (and forward) in time from the latest spin period measurement. 
The only assumption is the history of $L_X$ (i.e. $\dot{M}$).
Based on the derived spectral properties we argue that after the SN 2010da event, $L_X$ (see Fig.~\ref{fig:lc}) only varied by a small factor (i.e. $<$3), and the large deviation in the observed flux is mainly caused by a change in absorption \citepalias[see also][]{2018MNRAS.476L..45C}.
The theoretical prediction can then be tested with the observed data. 
In Fig. \ref{fig:evolP} we compare the observed spin periods with the theoretical period evolution for a range of $L_x$ values in order to account for small variations of $\dot{M}$.
The time axis range and reference date were selected for visualization purposes and to properly spread the observational points.

Assuming almost steady $\dot{M}$ after the SN 2010da event, the only way to explain the spin-period evolution is that an accretion disk was formed rotating retrograde with respect to the spin of the NS. 
Thus, the NS was initially spinning down and at some point the rotation was stopped; after that the NS started to rotate in the opposite direction (Fig.~\ref{fig:evolP}). 
In other words, from an evolutionary point of view, the retrograde disk appears to be a necessary initial condition in order to explain its spin evolution. While this raises questions of how the counter rotation was produced in the first place.
The formation of retrograde accretion disks has also been proposed to explain the long-term evolution of spin periods in X-ray binaries \citep{1997ApJ...488L.117N,2017arXiv170406364C}.
However, for the time window 2010-2015 we can only speculate for the evolution of $\dot{M}$ due to lack of observational coverage and because of the high absorption of the X-ray spectrum \citepalias[see Fig. 3 of][]{2018MNRAS.476L..45C}. 
Nevertheless, the $P$ and $\dot{P}$ measurements during the 2014 \cxo observation are consistent with the constant $\dot{M}$ assumption (see Fig. \ref{fig:PPdot_B}), supporting our finding that the decrease of the observed flux between 2010-2015 is only due to high absorption.
If $\dot{M}$ during that period was two orders of magnitude lower than after 2015 then it is possible that there was no NS spin reversal.
In this case the NS spin period should have been larger than a few hundred seconds prior to the SN 2010da event.

In general there are multiple ways to maintain the necessary $\dot{M}$ to fuel ULXPs.
For systems fueled by Roche Lobe (RL) overflow,  the accretion flow carries the angular momentum of the orbit.
Thus, it is implausible for an accretion disk to switch between retrograde and prograde rotation \citep{1997ApJ...481L.101C},
and any change between retrograde and prograde rotation of the NS disk system needs to be due to a change in the direction of the NS rotation.
Thus, for \ulx it is more plausible that the accretion mechanism was drastically altered during the SN 2010da event, prior of which the NS could have been fueled via wind accretion. 
A possible explanation for such a phase transition to occur can be an episode of thermal-timescale mass transfer \citep{2001ApJ...552L.109K}.
In that case the donor star has a radiative envelope and fills its RL as it expands across the Hertzsprung gap after the wind-fed X-ray phase ends.

However, RL overflow is not the only way to explain such high accretion rates. Supergiant systems and luminous blue variables (LBV) can also exhibit mass loss rates in excess of 10$^{-6}$ M$_\sun$ yr$^{-1}$ \citep{2014ARA&A..52..487S}, thus for a binary system in a moderately close orbit the compact object can capture enough matter to sustain super-Eddington accretion. 
An LBV is also most probable according to \citet{2016MNRAS.457.1636B} while other authors have ruled out this classification as inconclusive \citep{2016ApJ...830..142L,2016ApJ...830...11V}. 
We note that many of the arguments used in the classification of the system were based on the decay of the X-ray flux observed after the SN 2010da event. 
However, the decay in the X-ray flux was most likely a result of rapid increase in X-ray absorption \citepalias{2018MNRAS.476L..45C}.

\section{Conclusions}

We have studied the spin period evolution of \ulx, the most extreme spin-up NS powered by an accretion disk.
We derived precise measurements of $P$ and $\dot{P}$ for different epochs, and we have applied theoretical models of accretion theory in order to estimate the magnetic field strength of the NS.
Even though many of the models assume a geometrically thin disk we find that in the current regime of the system ($\omega_{\rm fast}<<1$) all of them can explain its evolution fairly well. 
Given the fast spin up of the NS, future observations when the spin period approaches its equilibrium will be of most importance to test different accretion models. 
However, given the time scale of the period evolution of the NS at the current accretion rate it might take close to one hundred years for the system to reach equilibrium. 
Finally, we argued that according to theoretical predictions and given the observable properties of the system, the NS has probably exhibited a spin reversal prior to 2014.

\bibliographystyle{aa} 
\bibliography{references} 

\begin{acknowledgements}
We would like to thank the editor S. Campana and the anonymous referee for their comments and suggestions that helped us improve this manuscript. 
We acknowledge the use of public data from \swift, \xmm, \cxo and \nus observatories.
We thank the Chandra director Dr. Belinda Wilkes for accepting our DDT request and the CXC team for promptly scheduling the observations.
This research has made use of software provided by the \cxo X-ray Center (CXC) in the application package CIAO.
We acknowledge the use of the \xmm SAS software, developed by ESA's Science Operations Centre staff.
\end{acknowledgements}

\begin{appendix} 

\section{Spectral analysis of \ulx}
\label{sec:spec}

\citetalias{2018MNRAS.476L..45C} have shown that the X-ray spectrum of \ulx can be described by a partially absorbed power-law and a disk black-body component. By comparing the 2010 and the 2016 \xmm spectra of the system, the authors found that the intrinsic spectrum has not significantly changed and the difference in the observed flux was only due to the heavily absorbed 2010 spectrum. 
Nevertheless, any model that is composed by a combination of a soft component and a hard (pulsating) component with  an exponential cutoff can explain the intrinsic X-ray spectrum. A partial covering absorber should be used to account for the long-term changes of the surrounding material in the binary system \citepalias[][]{2018MNRAS.476L..45C}.

In the current work the X-ray spectrum of \ulx was treated as follows; to model the soft component, we chose a simple disk black-body ({\tt diskbb} in {\tt xspec}), 
for the hard component we used a power-law ({\tt powerlaw}) modified by a high energy cut-off ({\tt highecut}).
The continuum is absorbed by interstellar material (modeled with {\tt tbabs}) and a partial covering absorber ({\tt pcfabs}) that accounts for intrinsic absorption. 
To mitigate the issues posed by limited statistics we adopted a Bayesian framework to fit the available X-ray spectra \citep[see also;][]{2018arXiv181106251K}. 
We used the Bayesian X-ray analysis package \citep[BXA][]{2014A&A...564A.125B} that was specifically 
designed to test various models in low-statistic spectra. 
In practice, BXA connects the nested sampling \citep{2004AIPC..735..395S} algorithm MultiNest \citep{2009MNRAS.398.1601F} with {\tt xspec}. BXA explores the parameter space and can be used for parameter estimation (probability distributions of each model parameter and their degeneracies).
We use BXA with its default parameters (400 live points and $\log Z$/evidence accuracy of 0.1). 

By using this Bayesian framework we are able to fit the phenomenological model described above to the X-ray spectrum of \ulx and derive a probability density function for $L_X$ of the system.  
Our results probe $L_X$ into three different periods; 
(I) the first 10 days after the SN 2010da event, where $L_X$ remained almost constant and the $N_H$ and covering fraction of the partial covering component rapidly increased,
(II) the following period till 2015, where only a couple of observations were performed and the X-ray spectrum is heavily absorbed ($N_H>$\ohcm{24}) making any $L_X$ estimation uncertain, and
(III) the period between 2016 and today, where the total $N_H$ dropped below \ohcm{22} and the inclusion of the partial coverage component is not statistically significant in the spectra (i.e. its parameters are not constrained). 
Our results suggest that \ulx maintained an almost constant $L_X$ (factor of 3) within the eight year period (2010--2018),
except one measurement in 2014, where due to limited statistics we cannot simultaneously constrain both $L_X$ and the partial absorber parameters.
In any case the upper limit for the $L_X$ of the system (as denoted by the 95\% percentile of the $L_X$ distribution) is only a factor of 3 lower than the $L_X$ derived from the 2016 \xmm observations.

\section{Temporal properties of \ulx: Accelerated Epoch Folding}
\label{sec:temp}

\begin{table*}[ht]
\caption{Properties of \ulx}
  \scalebox{0.9}{
 \begin{threeparttable}
\begin{tabular}{lccccccc}
\hline\noalign{\smallskip}
     \multicolumn{1}{c}{Target} &
     \multicolumn{1}{c}{Observatory/} &
     \multicolumn{1}{c}{T$_{\rm obs}$ $^{(b)}$} &
     \multicolumn{1}{c}{T$_{\rm Zero}$ $^{(c)}$} &
     \multicolumn{1}{c}{P$_{\rm Zero}$ $^{(d)}$} &  
     \multicolumn{1}{c}{$Log(|\dot{P}|)$ $^{(e)}$} &
     \multicolumn{1}{c}{PF $^{(f)}$} &
     \multicolumn{1}{c}{$\log(F_x)$ $^{(g)}$} \\
     
     \multicolumn{1}{c}{} &
     \multicolumn{1}{c}{ObsID $^{(a)}$} &
     \multicolumn{1}{c}{MJD} &
     \multicolumn{1}{c}{MJD--T$_{\rm obs}$} &
     \multicolumn{1}{c}{s} &   
     \multicolumn{1}{c}{s/s} &
     \multicolumn{1}{c}{} &
     \multicolumn{1}{c}{erg s$^{-1}$ cm$^{-2}$} \\

     \noalign{\smallskip}\hline\noalign{\smallskip}

\hline\noalign{\smallskip}
NGC300 ULX1 & S-00031726001 & 55341 & --& --& --& --& $-11.37^{+0.12}_{-0.09}$\\
        \noalign{\smallskip} 
& X-0656780401 & 55344.6 & --& --& --& --& $-11.60^{+0.24}_{-0.22}$\\
       \noalign{\smallskip}
& C-12238& 55463.1 &   --     &    --      & --    &  --  & $-12.86^{+0.7}_{-0.13}$\\
         \noalign{\smallskip}         
& C-16029& 56978.6 &   --     &    126.28$\pm$0.3$^{(*)}$       & --4.94$\pm$0.20$^{(*)}$    &  0.64$\pm$0.154  & $-12.0^{+0.3}_{-0.5}$\\
         \noalign{\smallskip}  
& S-00049834002 & 57493 & --& --& --& --& $-11.50^{+0.54}_{-0.24}$\\
        \noalign{\smallskip}          
& S-00049834005 &  57502         &  --    &  44.18$\pm$0.1$^{(*)}$    & --6.0$\pm$0.2$^{(*)}$    & 0.75$\pm$0.2& $-11.48^{+0.27}_{-0.10}$\\
         \noalign{\smallskip}
& N-30202035002 & 57738 &  0.65661582  & 31.718 $\pm$0.001     & --6.257$\pm$0.005     & 0.746$\pm$0.008  & --  \\
        \noalign{\smallskip}         
& X-0791010101 & 57739 &  0.39833581 & 31.683$\pm$0.001  & --6.25$\pm$0.01  & 0.607$\pm$0.005 & $-11.362^{+0.015}_{-0.013}$\\
        \noalign{\smallskip}
& X-0791010301 & 57741 &  0.39280867 & 31.588$\pm$0.001  & --6.25$\pm$0.01  & 0.613$\pm$0.007 & $-11.337^{+0.024}_{-0.017}$\\
        \noalign{\smallskip}  
& S-00049834008 & 57860        &  --   &  26.87$\pm$0.1$^{(*)}$ & --6.34$\pm$0.4$^{(*)}$    & 0.61$\pm$0.12& $-11.15^{+0.19}_{-0.09}$\\
         \noalign{\smallskip}
& S-00049834010 & 57866         &  --    &  26.65$\pm$0.1$^{(*)}$ &  --6.4$\pm$0.4$^{(*)}$     &  0.64$\pm$0.07 & $-11.21^{+0.21}_{-0.06}$\\
        \noalign{\smallskip}
& S-00049834013 & 57941        & --     &  24.24$\pm$0.1$^{(*)}$ & --6.8$\pm$0.4$^{(*)}$      &  0.7$\pm$0.12    & $-11.33^{+0.26}_{-0.11}$\\
        \noalign{\smallskip}
& S-00049834014 & 57946         &  --    &  24.22$\pm$0.1$^{(*)}$ &  --6.5$\pm$0.4$^{(*)}$      & 0.79$\pm$0.2   & $-11.37^{+0.25}_{-0.12}$\\
        \noalign{\smallskip}
& S-00049834015 & 58143         &  --    &  20.06$\pm$0.1$^{(*)}$ &  --6.5$\pm$0.4$^{(*)}$      & 0.79$\pm$0.2   & $-11.49^{+0.23}_{-0.08}$ \\ 
        \noalign{\smallskip}
& N-90401005002 & 58149 &  0.05840888  &  19.976$\pm$0.002 & --6.74$\pm$0.02    & 0.66$\pm$0.02 & -- \\
        \noalign{\smallskip}        
& C-20965& 58157 & 0.12920876 & 19.857$\pm$0.002& $<-6.7$ & 0.69$\pm$0.03 & $-11.38^{+0.18}_{-0.03}$ \\
         \noalign{\smallskip}
& C-20966& 58160 & 0.69721163 & 19.808$\pm$0.002&  $<-6.7$ & 0.66$\pm$0.03 & $-11.44^{+0.17}_{-0.03}$\\
         \noalign{\smallskip}
& C-20965/20966& 58157 & 0.12920876 & 19.857$\pm$0.002&  --6.82$\pm$0.02 & 0.68$\pm$0.02 & -- \\
         \noalign{\smallskip}
& S-00049834019 & 58221         &  --    &  19.046$\pm$0.01$^{(*)}$ &  --6.6$\pm$0.6$^{(*)}$      & 0.75$\pm$0.1   & $-11.5^{+0.4}_{-0.1}$ \\ 
        \noalign{\smallskip}           
\hline\noalign{\smallskip}
\hline\noalign{\smallskip}
\end{tabular}
\footnotesize
\tnote{(a)} Observation ID for \xmm (X), \cxo (C) and \swift (S).
\tnote{(b)} T$_{\rm obs}$ : Start day of observation.
\tnote{(c)} T$_{\rm Zero}$ : reference time for calculated P$_{\rm Zero}$ and $\dot{P}$ using millisecond accuracy.
\tnote{(d)} Pulse period derived using the accelerated epoch folding algorithm.
\tnote{(e)} Spin-up of the pulsar. Data sets with low statistics are marked with star (*), for those a unique solution could not be established thus the given uncertainty denotes the characteristic spread between multiple solutions. 
\tnote{(f)} Pulsed fraction ($PF=(F_{max}-F_{min})/(F_{max}+F_{min})$) derived form the folded pulse profile using 10 phase bins, we treated uncertainties in phase bins with low numbers of counts by following \citet{1986ApJ...303..336G}.
\tnote{(g)} Absorption corrected X-ray Flux (0.3--10.0 keV) of the hard component as derived from the distributions of the marginalised parameters using the Bayesian framework described in the text. Given value denotes the median of the distribution, while error values the 5\% and 95\% percentiles. 
\end{threeparttable}
 }
\label{tab:log}
\end{table*}

Our aim is to study the temporal properties of \ulx on short time scales. Thus we want to determine both $P$ and $\dot{P}$ from a single long observation, or by using multiple short exposures (i.e. snapshots) performed within a couple of days. 
We extracted the barycentric corrected event arrival times and performed various statistical tests to search for a periodic signal. 
For individual observations with a high number of events (i.e. \xmm, \nus), we performed an accelerated epoch folding (AEF) test \citep{1983ApJ...272..256L} to derive the ephemeris of the pulsar similar to the work of \citet{2016ApJ...831L..14F}.
The AEF method uses a grid of trial points in the $P$-$\dot{P}$ plane in order to fold a time series and create a pulse profile, and performs a $\chi^2$ test based on the constant signal hypothesis.
Thus, a maximization of the $\chi^2$ indirectly supports the presence of a periodic signal.
The significance of any detection is then assessed by MCMC simulations. 
Fake datasets can be created and analyzed by the AEF algorithm, the results can then be used to estimate a baseline for the maximum $\chi^2$ value we would expect from a non periodic time series.

For each observation of \ulx we used a grid of points in the $P$-$\dot{P}$ space to determine the best fit value. We then performed multiple iterations by decreasing the bin size of the grid around the best fit values until we scanned the $P$-$\dot{P}$ plane with sufficient resolution. 
For the 2016 \xmm and \nus (2016 and 2018) datasets we used a grid of equally spaced points, in linear space for $P$ ($\sim$0.5-1 ms resolution), and in logarithmic space for $\dot{P}$ ($\sim$0.002-0.01 resolution). 
The final grid was smoothed using the four closest neighbors to avoid numerical artifacts.
To calculate the uncertainties of the best fit values we introduced an additional step, for each grid point we calculated the pulsed fraction (PF) of the folded pulse profile. In a statistical sense, the uncertainties of the best fit values in the $P$-$\dot{P}$ can be derived as the PF diverges from the best fit value by a factor proportional to the uncertainty (statistical plus 5\% systematic) of the pulsed fraction (see Figs. \ref{fig:ppdot} \& \ref{fig:ppdot2}).
We note that for observations with good statistics this estimation is insensitive to the selected phase-bin size for the pulse profile as long as the number of counts within each phase bin obey Gaussian statistics; For \ulx (\xmm obsid:0791010101) the PF is 0.603$\pm$0.007 and 0.638$\pm$0.014 using 10 and 50 phase bins respectively.

The AEF method can be successfully applied to the two 2018 \cxo observations (obs-ids: 20965 \& 20966) that were performed within four days.
By analyzing each observation separately we can derive an accurate period and an upper limit for the $\dot{P}$ (see Table \ref{tab:log}).
By combining the events we can measure both $P$ and $\dot{P}$ (see Fig. \ref{fig:ppdot3}).

By using the AEF method in the 2014 \cxo observation and the \swift/XRT observations performed after 2015 we were also able to measure $P$ and $\dot{P}$.
However, these data sets suffer from low statistics thus any period search results in large uncertainties. 
Moreover, \swift data were collected from numerous snapshots over one to two days, thus the derived 2D ($P$-$\dot{P}$ plane) periodograms show multiple solutions. 
This degeneracy is due to the fact that the uncertainty of the derived spin period of each \swift/XRT snapshot is comparable to the spin period change between snapshots.
To mitigate these effects we need to use multiple \swift/XRT exposures that span over a duration of days to a week (see Fig. \ref{fig:ppdot4}).
Unfortunately, most of the \swift/XRT observations performed prior to the discovery of the spin period of \ulx are not suitable for detailed temporal studies as they were composed by snapshots with short exposures that were performed within $\sim$24 h.
Thus, we can only compute ${\Delta}P$ for large time intervals (i.e. months-years) and no secular $\dot{P}$ for weekly intervals. 
The results of our analysis are summarized in Table \ref{tab:log}. 
For some datasets with gaps and low statistics we could not determine a unique set of values due to aliasing and multiple possible solutions (see values marked with ``*'' in Table \ref{tab:log}).

At this point we make a note about the ongoing and future monitoring X-ray observations of \ulx with \swift/XRT that will be presented in a future study.
These monitoring observations are repeated with a cadence of 3-4 days.
Thus, combining multiple observations performed within a few days we are able to constrain both $P$ and $\dot{P}$ using the AEF method. 
An example of our analysis of these datasets is shown in the right panel of Fig. \ref{fig:ppdot4}.  

\begin{figure*}
  \resizebox{\hsize}{!}{
  \includegraphics[angle=0,clip=]{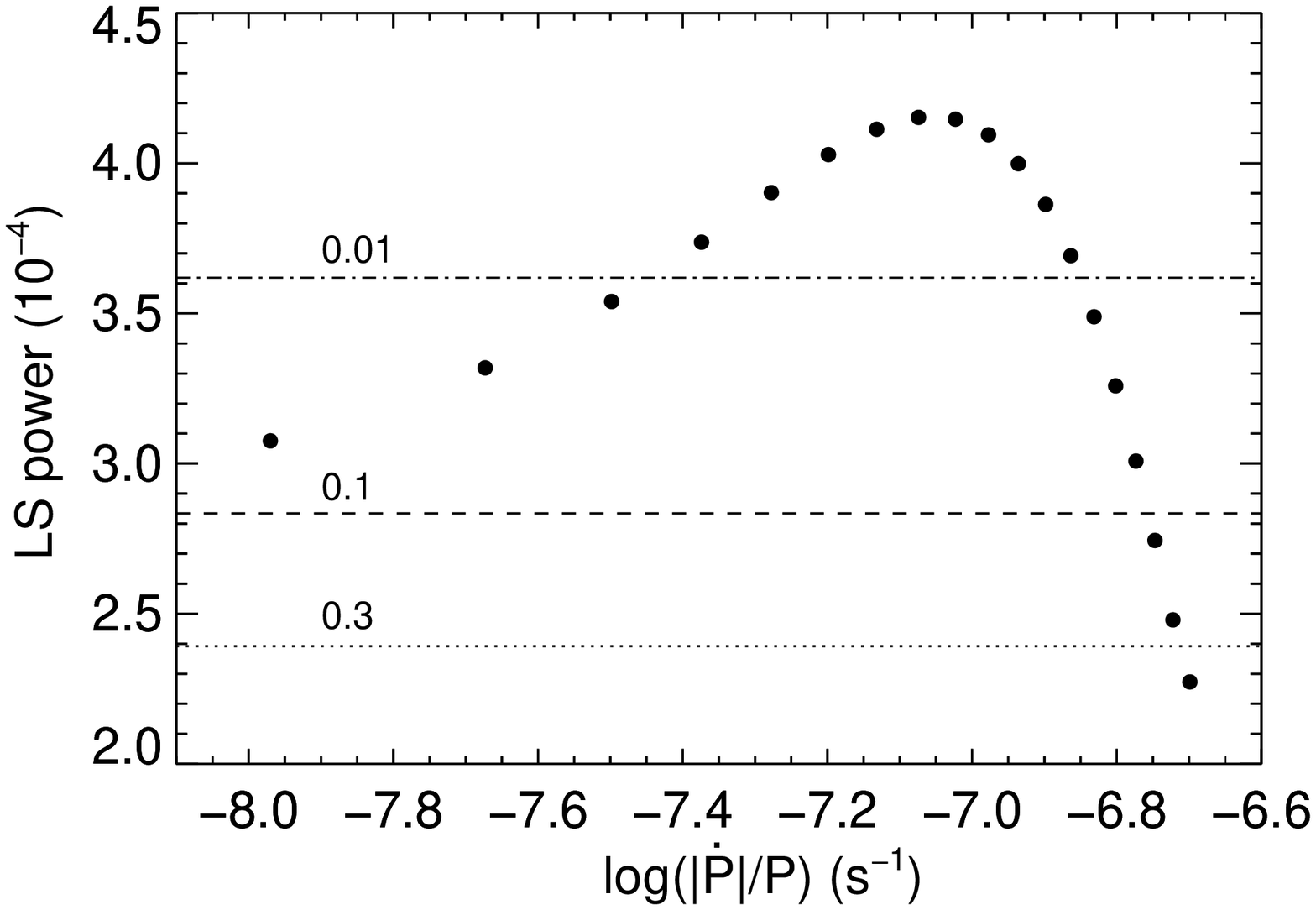}
  \includegraphics[angle=0,clip=]{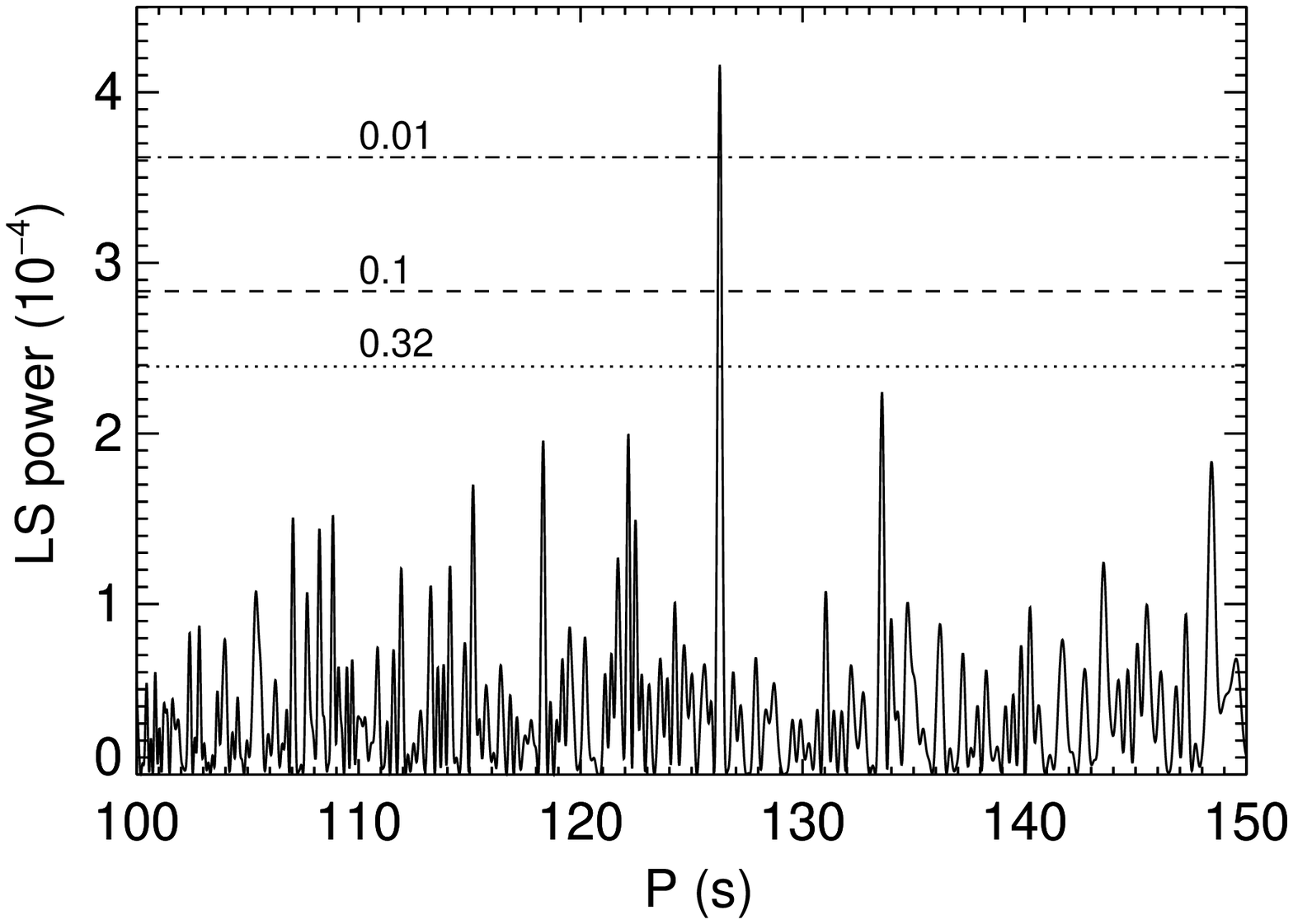}
 }
  \caption{\emph{Left:} Maximum power of the most significant peak of the periodogram as derived for 20 time series that were stretched for different $\dot{P}/P$ values for the \cxo observation performed in Nov. 2014. \emph{Right:} Periodogram for the most significant detection with $\dot{P}/P=-8.84\times10^{-8} s^{-1}$. Horizontal lines correspond to the 32\%, 10\% and 1\% false alarm probability.}
  \label{fig:pdotp_max}
\end{figure*}

To asses the significance of the observations with low statistics we performed MCMC simulations.
We created simulated datasets with similar observational characteristics as the \swift/XRT or \cxo data sets (i.e. number of counts and good time intervals). 
For each original dataset we performed 10$^{6-7}$ tests by creating 100-1000 fake datasets that were analyzed by our AEF algorithm.
For all cases, we determined that for the fake datasets the probability of obtaining $\chi^2$ values as high as the ones measured in the observed datasets is smaller than $\sim$10$^{-6}$ (or 4.8$\sigma$).

To further investigate the significance of the detected periods we computed the Lomb-Scargle (LS) periodogram \citep{1983ApJ...266..160L} of the de-accelerated event arrival times. As the recorded arrival times are a result of almost constant acceleration we can transform the time scale during the observation so that the pulsar period will remain constant. The transformation is based on a simple Taylor expansion and can be given by equation:
\begin{equation}
t'=t+\frac{t^2\dot{P}}{2P},\nonumber
\end{equation}
where t is the detected event time and t' the transformed. For each observation with low statistics we calculated the LS periodogram for a series of $\dot{P}/P$ values while performing white-noise simulations to derive the significance of the detected periods. In Fig.\,\ref{fig:pdotp_max} we show an example of our test for the \cxo observation (obsid: 16029) where the 126~s period was discovered.
To estimate the significance of the periodic signal as computed by the LS peridogram we used the block
bootstrap method \citep{10.2307/3182810}.
Time series was re-sampled with replacement within time blocks of 20~s, while the starting points of the blocks where shuffled.
The significance of the periodic signal was then estimated by
simulating 10000 light curves \citep[e.g. see also][]{2017A&A...602A..81C}.

For completeness we note the following. \ulx was not active between 2000 and 2005 as it was not detected during four deep ($\sim$40 ks) \xmm observations (obs-ids: 0112800101, 0112800201, 0305860301, 0305860401).
Moreover, of particular interest is the 2010 \xmm observation (obsid: 0656780401) with duration of 18 ks.
We performed periodicity tests between 0.1 and 5000 s using a grid of $\dot{P}/P$ values, but
we could not confirm the presence of any periodic modulation.

To conclude, from our analysis we were able to measure the spin evolution of \ulx over a period of about four years (2014-2018), prior to which no spin measurement could be determined.
We were also able to determine the instantaneous $P$ and $\dot{P}$ values using the \xmm and \nus 2016 data and the \cxo 2018 data.
For all other observations the derived spin periods should be considered as averaged  within the exposure duration. 
Consequently, we can use these values to determine the long-term (secular) spin-up rate of the pulsar, but cannot be considered as derivatives of $P$ at a specific time.

\begin{figure*}
  \resizebox{\hsize}{!}{
     \includegraphics[angle=0,clip=]{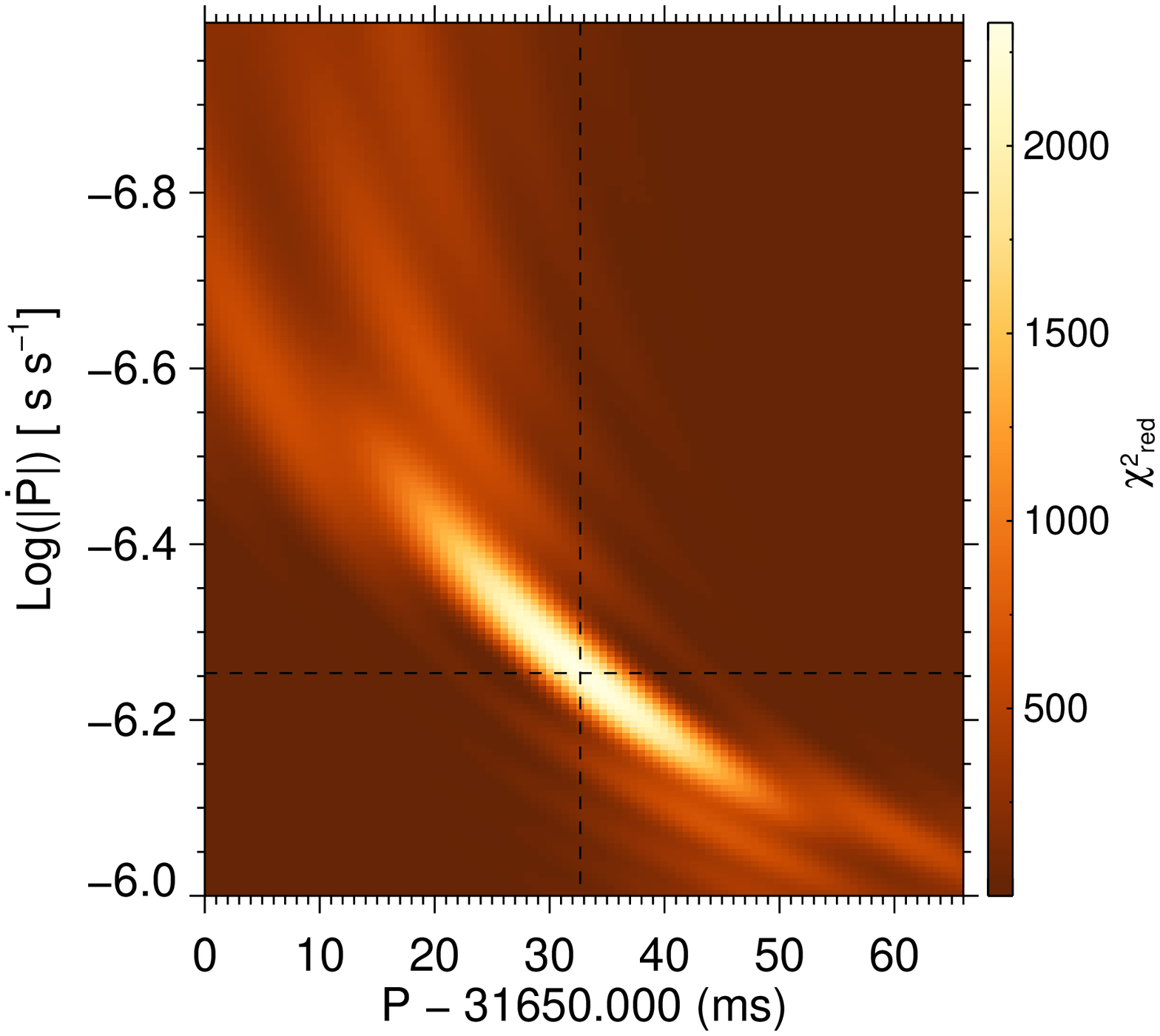}
     \includegraphics[angle=0,clip=]{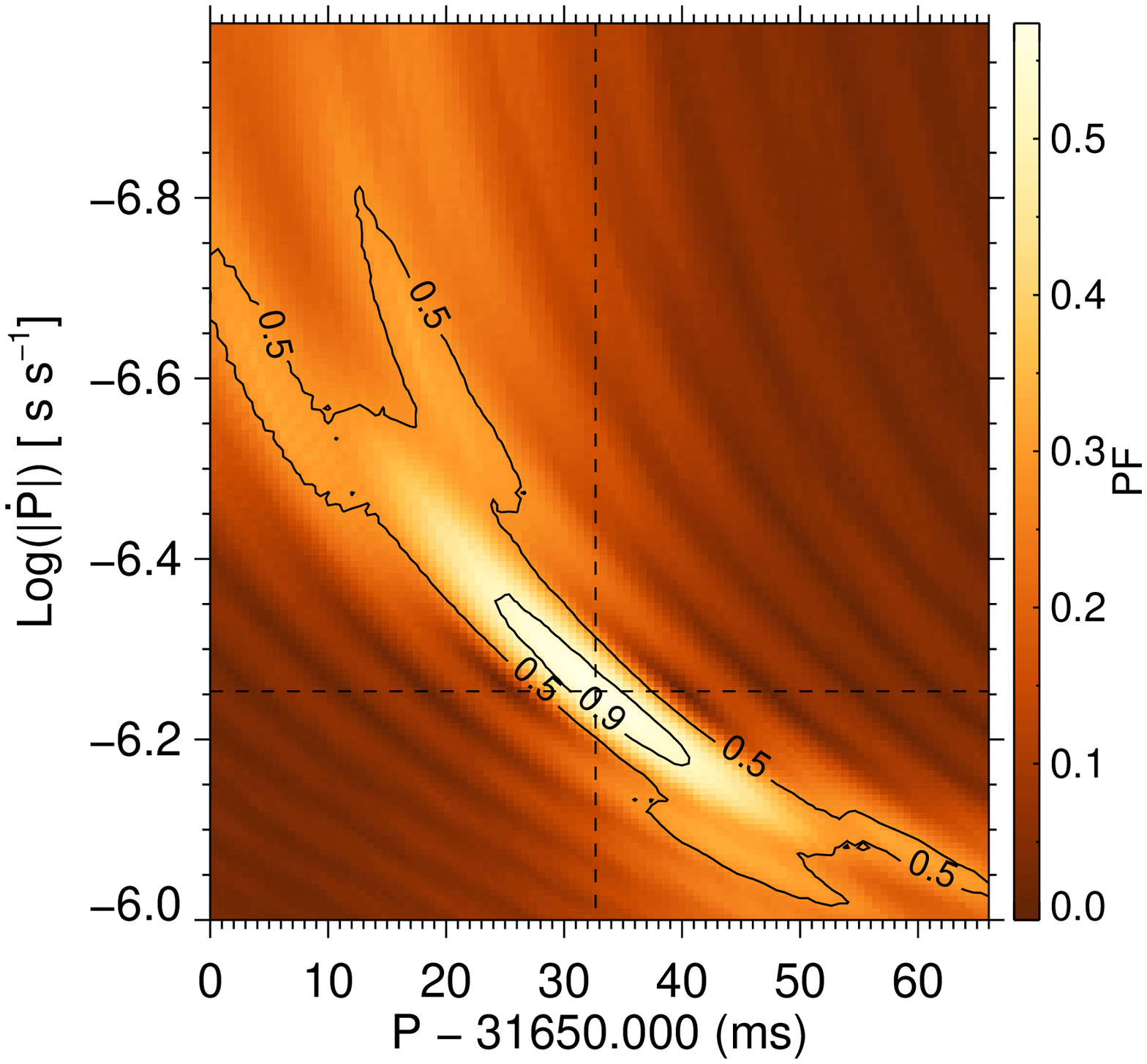}}
  \caption{{\emph{Left:}} Results of the accelerated search in the 2016 \xmm data of \ulx. {\emph{Right:}} Computed pulsed fraction from the corresponding pulse profiles using 50 phase bins. Contours mark the regions where the pulse profile has decreased by 10\% and 50\% of its maximum value.} 
  \label{fig:ppdot}
\end{figure*}

\begin{figure*}
    \resizebox{\hsize}{!}{
     \includegraphics[angle=0,clip=]{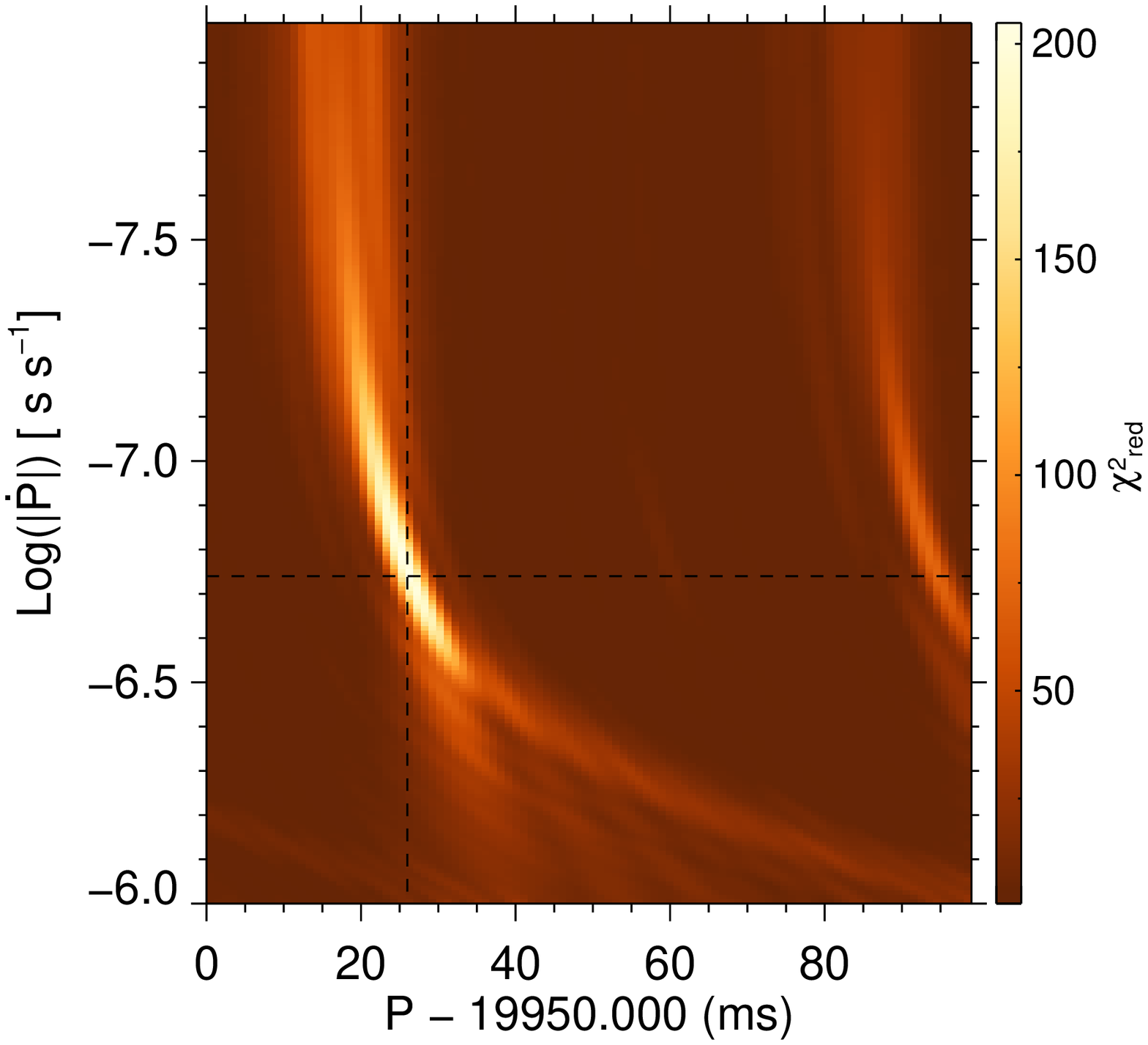}
     \includegraphics[angle=0,clip=]{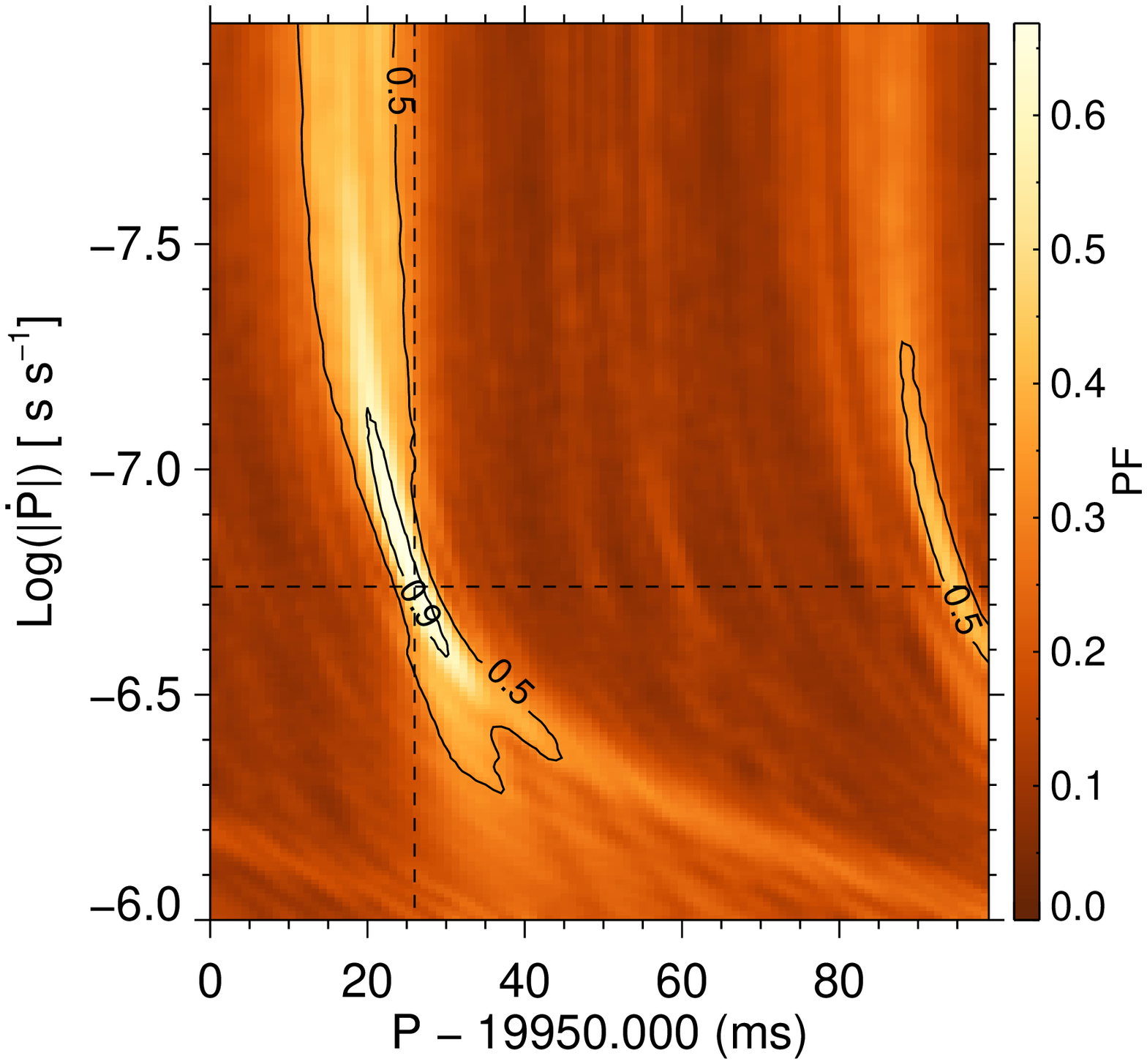}}   
  \caption{Same as Fig. \ref{fig:ppdot}, but for the 2018 \nus data (obsid 90401005002). We note the appearance of ``islands'' of multiple solutions, but due to the good statistics we are able to avoid any degeneracy. The goodness of fit (i.e. $\chi^2$) is computed by using 10 phase bins for the folded pulse profile.} 
  \label{fig:ppdot2}
\end{figure*}

\begin{figure*}
  \resizebox{\hsize}{!}{
     \includegraphics[angle=0,clip=]{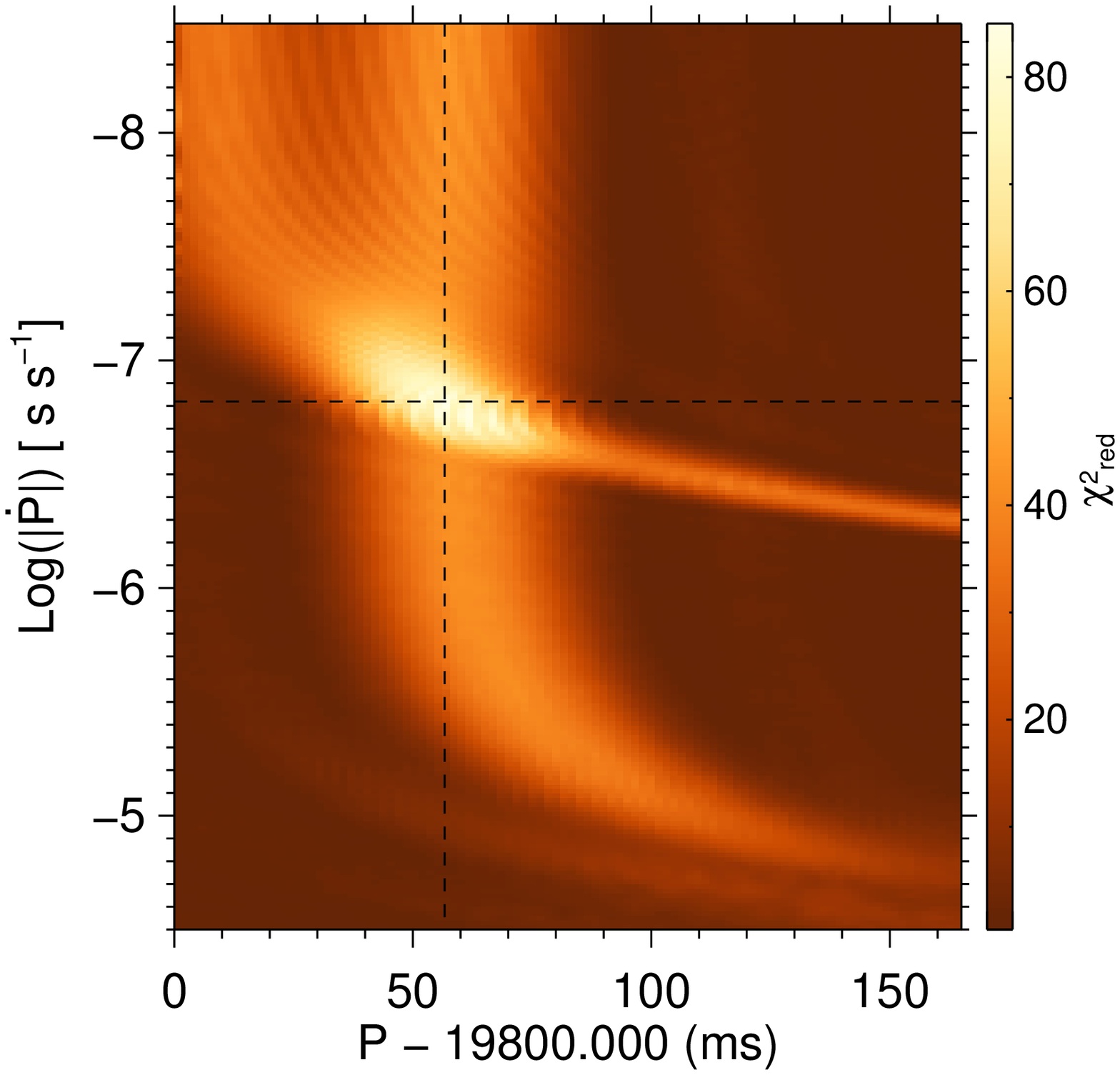}
     \includegraphics[angle=0,clip=]{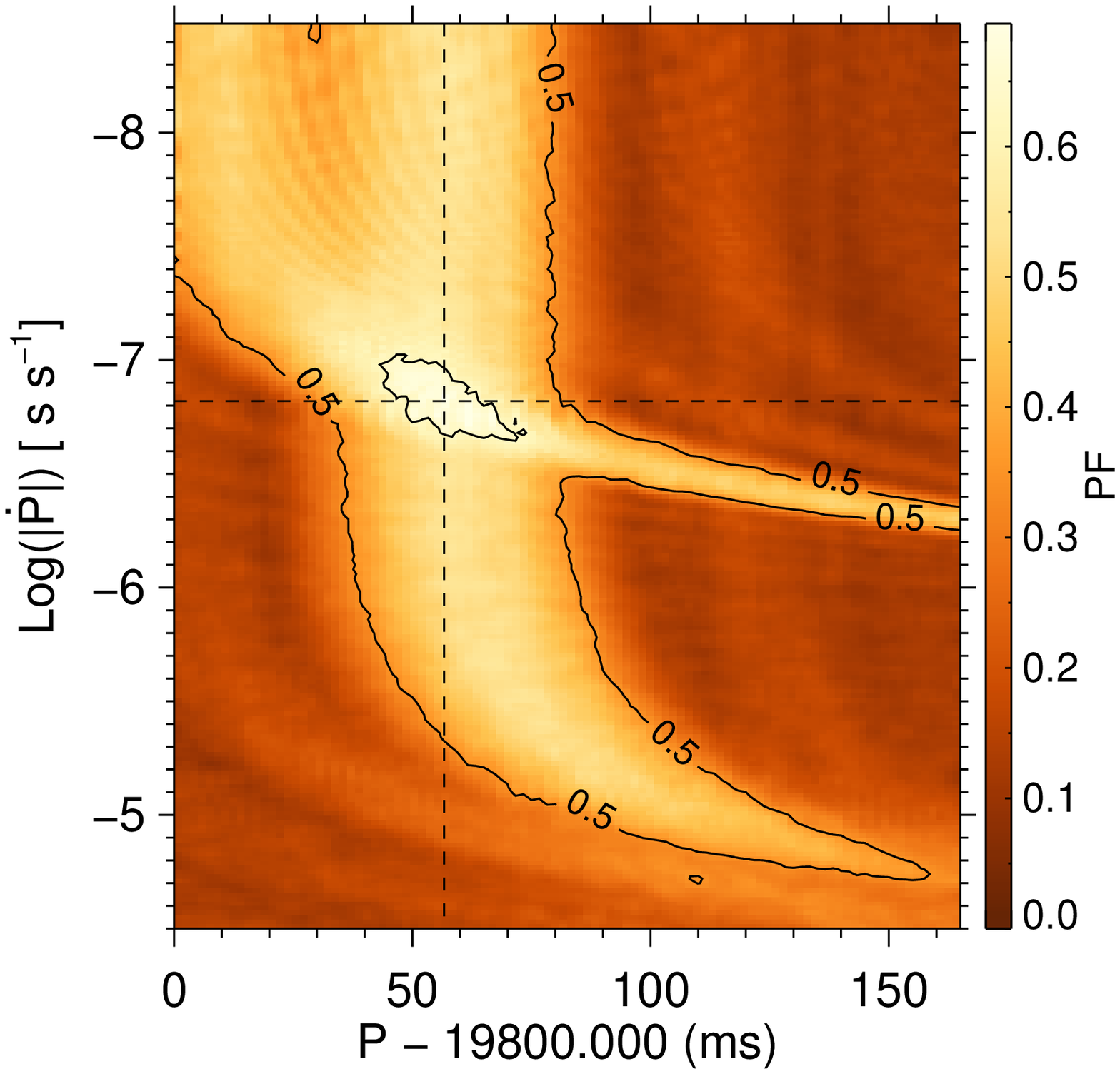}}
  \caption{Same as Fig. \ref{fig:ppdot}, but for the combined \cxo data obtained from  MJD 58157 to 58180. We note the difference in the 50\% contour of the pulsed fraction (and the $\chi^2$) compared to Fig. \ref{fig:ppdot}. This is an effect of the time gap between the observations. In the case of \swift/XRT due to the observational gaps and the low number of counts it is not possible to derive a unique solution in the $P$ vs. $\dot{P}$ plane. The goodness of fit (i.e. $\chi^2$) is computed by using 10 phase bins for the folded pulse profile.} 
  \label{fig:ppdot3}
\end{figure*}

\begin{figure*}
  \resizebox{\hsize}{!}{
     \includegraphics[angle=0,clip=]{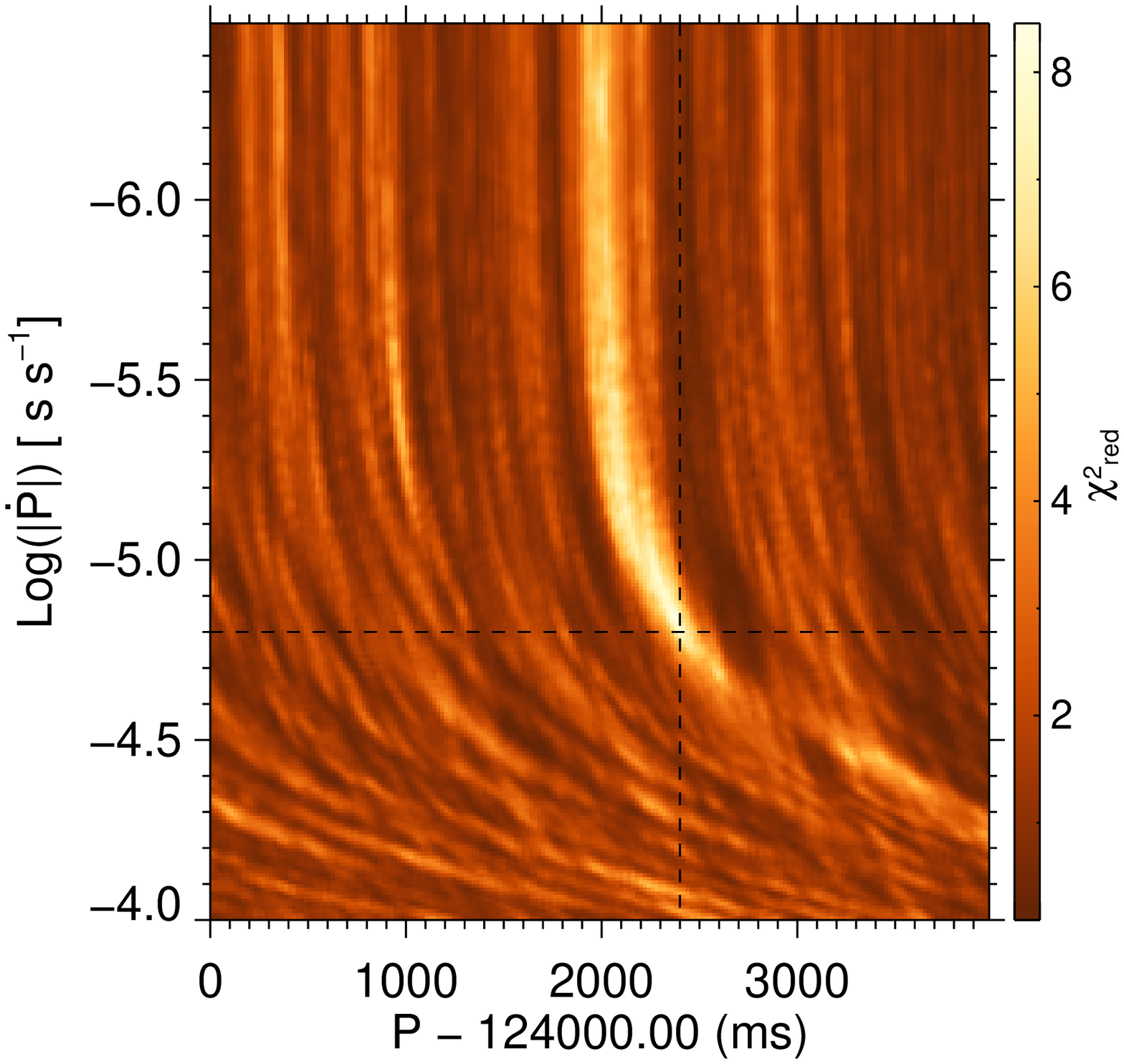}
     \includegraphics[angle=0,clip=]{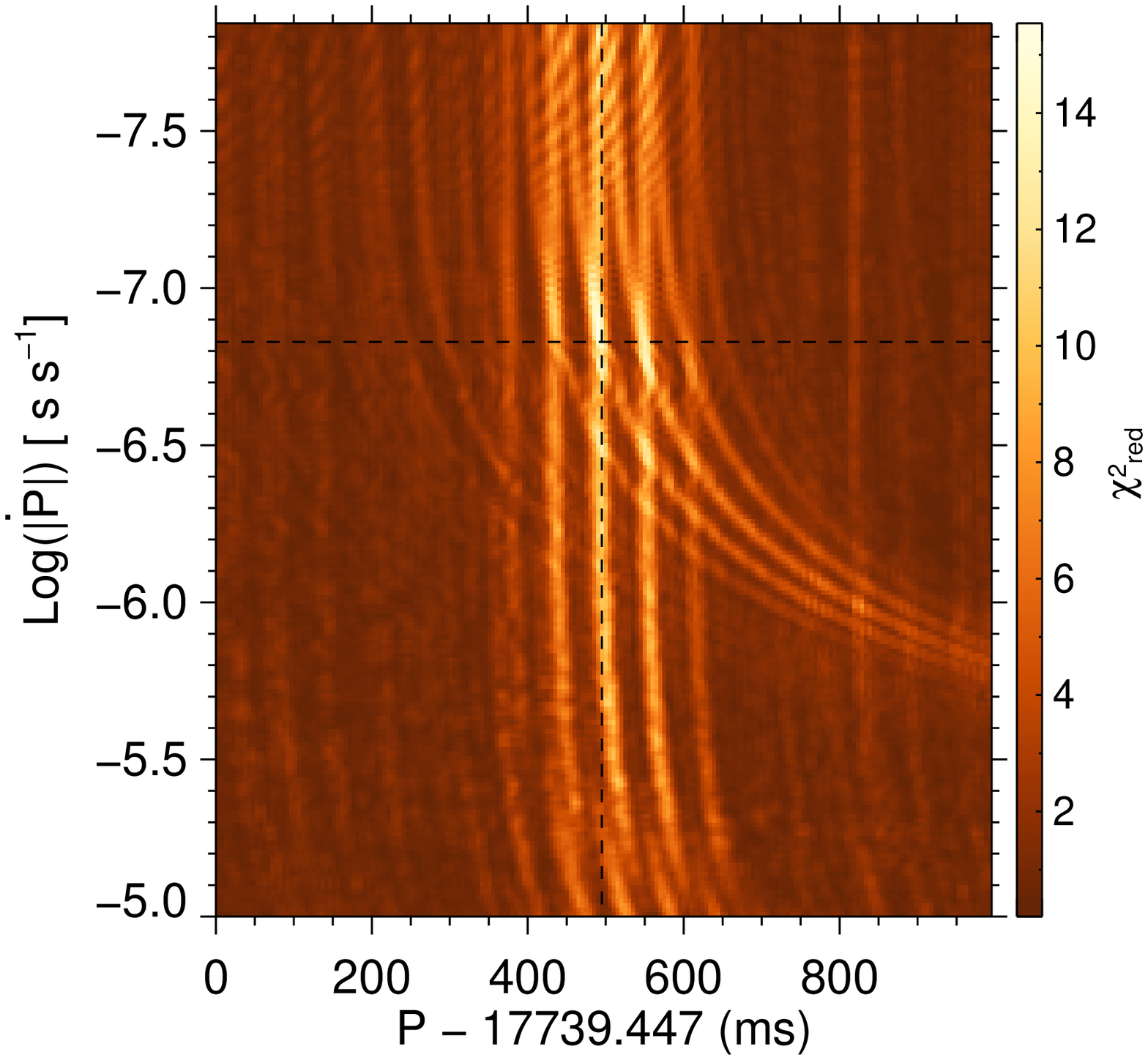}}
  \caption{Same as left panel of Fig. \ref{fig:ppdot}, but for the 2014 \cxo data (left) and two combined \swift/XRT observations (obsids: 00049834047-8) performed between MJD 58285-9 . Due to the low statistics only 5 phase bins were used for the AEF test.} 
  \label{fig:ppdot4}
\end{figure*}

\end{appendix}

\end{document}